\documentclass[a4paper,11pt]{article}

\usepackage{graphicx}  
\usepackage[final]{epsfig}
\usepackage{ifpdf}
\usepackage{amssymb,latexsym}
\usepackage[mathscr]{eucal}
\usepackage{url}
\usepackage{hyperref}
\usepackage{ntheorem-hyper}

\newcommand{\zz}{z-z_0}
\newcommand{\zzp}{(z-z_0)}
\newcommand{\ozzp}{(\oz-\oz_0)}

\newcommand{\varphiT}{\tilde \varphi}
\newcommand{\ovarphiT}{\overline{\tilde \varphi}}

\newcommand{\ii}{\mathrm{i}}

\newcommand{\hsone}{\mathring{ \sigma}_1}
\newcommand{\hstwo}{\mathring{ \sigma}_2}

\newcommand{\Ef}{E_{\mcE,\varphi}}

\newcommand{\oz}{{\bar z}}

\newcommand{\hE}E
\newcommand{\hR}R
\newcommand{\hm}m

\newcommand{\ogamma }{{\bar \gamma}}

\newcommand{\FS}       
                  {F}

\newcommand{\HS} 
       {H_{\mbox{\scriptsize volume}}}

\newcommand{\bgamma}{\bar \gamma}

{\ptc{this should be removed in the oberwolfach version}}%

\newcommand{\mcE}{{\mycal E}}%

\newcommand{\eeal}[1]{\label{#1}\end{eqnarray}}
\newcommand{\C}{{\mathbb C}}
\newcommand{\bed}{\begin{deqarr}}
\newcommand{\eed}{\end{deqarr}}
\newcommand{\bedl}[1]{\begin{deqarr}\label{#1}}
\newcommand{\eedl}[2]{\arrlabel{#1}\label{#2}\end{deqarr}}

\newcommand{\omcE}{\,\,\overline{\!\!\mcE}}

\newcommand{\bel}[1]{\begin{equation}\label{#1}}
\newcommand{\bea}{\begin{eqnarray}}
\newcommand{\bean}{\begin{eqnarray}\nonumber}
\newcommand{\beal}[1]{\begin{eqnarray}\label{#1}}
\newcommand{\eea}{\end{eqnarray}}

\newcommand{\Eq}[1]{Equation~\eq{#1}}

\newcommand{\qed}{\hfill $\Box$ \medskip}
\newcommand{\proof}{\noindent {\sc Proof:\ }}
\newcommand{\be}{\begin{equation}}
\newcommand{\eeq}{\end{equation}}
\newcommand{\ee}{\end{equation}}
\newcommand{\beqa}{\begin{eqnarray}}
\newcommand{\eeqa}{\end{eqnarray}}
\newcommand{\beqan}{\begin{eqnarray*}}
\newcommand{\eeqan}{\end{eqnarray*}}
\newcommand{\ba}{\begin{array}}
\newcommand{\ea}{\end{array}}

\newcommand{\mcM}{{\mycal M}}

\newtheorem{Theorem} {\sc  Theorem\rm} [section]
\newtheorem{Corollary} [Theorem] {\sc  Corollary\rm}

\newtheorem{Proposition} [Theorem] {\sc  Proposition\rm}

\theorembodyfont{\upshape}
\newtheorem{Remark}[Theorem]{\sc  Remark\rm}

\DeclareFontFamily{OT1}{rsfs}{}
\DeclareFontShape{OT1}{rsfs}{m}{n}{ <-7> rsfs5 <7-10> rsfs7 <10-> rsfs10}{}
\DeclareMathAlphabet{\mycal}{OT1}{rsfs}{m}{n}

{\catcode `\@=11 \global\let\AddToReset=\@addtoreset}
\AddToReset{equation}{section}

\newcounter{mnotecount}[section]

\renewcommand{\themnotecount}{\thesection.\arabic{mnotecount}}

\newcommand{\mnote}[1]
{\protect{\stepcounter{mnotecount}}$^{\mbox{\footnotesize
$
\bullet$\protect\themnotecount}}$ \marginpar{
\raggedright\tiny\em
$\!\!\!\!\!\!\,\bullet$\protect\themnotecount: #1} }

\newcommand{\warn}[1]
{\protect{\stepcounter{mnotecount}}$^{\mbox{\footnotesize
$
\bullet$\themnotecount}}$ \marginpar{
\raggedright\tiny\em
$\!\!\!\!\!\!\,\bullet$\protect\themnotecount: {\bf Warning:} #1} }

\newcommand{\R}{\mathbb R}

\newcommand{\eq}[1]{(\ref{#1})}

\newcommand{\ptc}[1]{\mnote{{\bf ptc:}#1}}

\newcommand{\beqar}{\begin{deqarr}}
\newcommand{\eeqar}{\end{deqarr}}

\newcommand{\beaa}{\begin{eqnarray*}}
\newcommand{\eeaa}{\end{eqnarray*}}

\newcommand{\ovphi}{\,\overline{\!\varphi}}

\begin{document}
\title{On the Ernst electro-vacuum equations\\ and ergosurfaces}
\date{}
\author{Piotr T.\
  Chru\'sciel\thanks{\href{http://www.phys.univ-tours.fr/\string~piotr}{\tt
      http://www.phys.univ-tours.fr/$\sim$piotr}, e-mail:
    \protect\href{mailto:chrusciel@maths.ox.ac.uk}{\tt chrusciel@maths.ox.ac.uk}}
   \vspace{0.18cm} \\
LMPT,
F\'ed\'eration Denis Poisson, Tours \\
Mathematical Institute and Hertford College, Oxford
  \\
\\ Sebastian J. Szybka\thanks{Partially supported within the framework
of the European Associated Laboratory ``Astrophysics Poland--France'' and by the MNII grant 1 P03B 012 29; { e-mail:}
  \mbox{\protect\href{mailto:szybka@if.uj.edu.pl}{\tt szybka@if.uj.edu.pl}}} 
\vspace{0.18cm}\\ 
Obserwatorium Astronomiczne, Centrum Astrofizyki
\\ Uniwersytet Jagiello\'nski, Krak\'ow}

\maketitle
\begin{abstract}
The question of smoothness at the ergosurface of the space-time
metric constructed out of solutions $(\mcE,\varphi)$ of the Ernst
electro-vacuum equations is considered.
We prove smoothness of those ergosurfaces  at which
$\Re \mcE$ provides the dominant contribution to $f=-(\Re \mcE +
|\varphi|^2)$ at the zero-level-set of $f$. Some partial results are
obtained in the remaining cases: in particular we give examples of
leading-order solutions with singular isolated ``ergocircles".
\end{abstract}

\section{Introduction}

In recent work~\cite{CGMS} we have shown that a vacuum space-time
metric is smooth near a ``Ernst ergosurface" $E_\mcE =\{\Re
\mcE=0\;, \ \rho \ne 0\}$ if and only if the Ernst potential $\mcE$
is smooth near $E_\mcE$ and does not have zeros of infinite order
there. It is of interest to enquire whether a similar property holds
for electro-vacuum metrics. While we have not been able to obtain a
complete answer to this question, in this note we present a series
of partial results, amongst which:

\begin{Theorem}
\label{TA}  Consider a smooth solution $( \mcE,\varphi)$ of the
electro-vacuum Ernst equations \eq{e1}-\eq{e2} below, and let the
Ernst ergosurface $\Ef$ be defined as the set
\bel{Ereg} \Ef :=\{\mcE+\omcE +2\ovphi\varphi=0\;, \ \rho \ne 0
\}\;.
 \ee
Suppose that $\mcE +\omcE$ has a zero of finite order at $\Ef$. If
the $\varphi$ terms contribute subleading terms to $\mcE+\omcE
+2\ovphi\varphi$ at $\Ef$, then there exists a neighborhood of $\Ef$
on which the tensor field \eq{2.1} obtained by solving
\eq{e4}-\eq{e5} is smooth and has Lorentzian signature.
\end{Theorem}

Theorem~\ref{TA} is proved in Section~\ref{SEdom}.

 To make things
clear, consider a point $p$ at which
$$
 f:=- \frac 12 (\mcE+\omcE
+2\ovphi\varphi)
$$
vanishes. Expanding $\mcE$ and $\varphi$ in a Taylor series at $p$,
let $m$ be the order of   the leading Taylor polynomial of $ \Re
\mcE-\Re\mcE(p)$, and let $k$ be the corresponding order for
$\varphi-\varphi(p)$. Then we say that  the $\varphi$ terms
contribute subleading terms to $f$ if $2k >m$.

Under the remaining conditions of Theorem~\ref{TA}, the condition of
a zero of finite order is \emph{necessary and sufficient}, as
smoothness of the metric near $\Ef$ implies analyticity of $\mcE$
and $\varphi$.

It follows from the analysis in \cite{CGMS} that, in vacuum, a
generic point on $\Ef$ will be a zero of $\mcE$ of order one. One
expects this result to remain true in electro-vacuum, so that
Theorem~\ref{TA} should cover generic situations.

A significant application of Theorem~\ref{TA}, to solutions obtained
by applying a Harrison transformation to a vacuum solution, is given
in Section~\ref{SHarrison} below.

 Some partial results, presented in Section~\ref{Spartial}, are
obtained in the cases not covered by Theorem~\ref{TA}:  We describe
completely the leading-order behavior of $\varphi$  at those
ergosurfaces at which $\varphi$ provides the dominant contribution
to $f$. We show that there exist Taylor polynomials solving the
Ernst equation at leading order which result in singularities of the
space-time metric on $\Ef$. This result does not, however, prove
that there exist smooth solutions of the electro-vacuum Ernst
equations which lead to metrics which are singular at the
ergosurface because it is not clear that the ``leading-order
solutions" that we construct correspond to solutions of the full,
non-truncated equations.

\section{Preliminaries}

 We use the same parameterisation of
the metric as in \cite{CGMS}:
\bel{2.1}
ds^2=f^{-1}\left[h\left(d\rho^2+d\zeta^2\right)+\rho^2d\phi^2\right]-f\left(d
t+ad\phi\right)^2\;,
 \ee
 with all functions depending only upon $\rho$ and $\zeta$.
 In electro-vacuum the Ernst equations
form a system of two coupled partial differential equations for two
complex valued functions $\mcE$ and $\varphi$
\cite{Exactsolutions2}, which we assume to be smooth:
\begin{eqnarray}
\label{e1} \left(\mcE+\omcE +2\ovphi\varphi\right)L\mcE
 &=&
 \left(\frac{\partial\mcE}{\partial\oz }+2\ovphi\frac{\partial\varphi}{\partial\oz }\right)
    \frac{\partial\mcE}{\partial
    z}
    +
     \left(\frac{\partial\mcE}{\partial
    z}+ 2 \ovphi\frac{\partial\varphi}{\partial
    z}\right)
    \frac{\partial\mcE}{\partial\oz }
    \;,
    \phantom{xxx}
    \\
\left(\mcE+\omcE +2\ovphi\varphi\right)L\varphi &=&
      \Big(\frac{\partial\mcE}{\partial\oz }+2\ovphi\frac{\partial\varphi}{\partial
    \oz}\Big)\frac{\partial\varphi}{\partial
    z}
    +
     \Big(\frac{\partial\mcE}{\partial
    z}+2\ovphi\frac{\partial\varphi}{\partial
    z}\Big)\frac{\partial\varphi}{\partial\oz }
    \;,
    \label{e2}
\end{eqnarray}
where
$$
 L=\frac{\partial^2}{\partial z\partial\overline
  z}+\frac{1}{2(z+\oz )}\Big(\frac{\partial}{\partial z}
  +\frac{\partial}{\partial\oz }\Big)
  \;,
  $$
with $z=\rho+\ii \zeta$.  The metric functions are determined from%
\footnote{Note that $\mcE$ here is minus $\mcE$ in \cite{CGMS}.}
\begin{eqnarray}
\label{e3}
 f &=& -\frac 12 (\mcE+\omcE +2\ovphi\varphi)
 \;,
 \\
 \label{e4}
  \frac{\partial
h}{\partial z}&=&
 (z+\oz )h\left(\frac{1}{2}{
 \displaystyle\left(\frac{\partial
\mcE}{\partial z}+2\ovphi\frac{
 \displaystyle\partial \varphi}{\partial
z}\right)\left(\frac{\partial\omcE }{\partial
z}+2\varphi\frac{\partial \ovphi}{\partial z}\right)}f^{-2}
 +{
 \displaystyle2\frac{\partial \ovphi}{\partial
      z}\frac{\partial \varphi}{\partial z}}f^{-1}\right),
      \phantom{xxxxxx}
      \\
\frac{\partial a}{\partial z}&=&
 \frac{1}{4}\displaystyle
 (z+\overline
z)\left({
 \displaystyle \frac{\partial\mcE}{\partial z}+2\ovphi\frac{\partial \varphi}{\partial z}-\frac{
 \displaystyle\partial\omcE }{\partial z}
 -2\varphi\frac{\partial
  \ovphi}{\partial z}}\right)f^{-2}
  \;.
  \label{e5}
\end{eqnarray}
The equations are singular at the \emph{Ernst ergosurface} $\Ef$
defined by \eq{Ereg}.

Let $\lambda\in \C$, $\mu \in \R$, then the following transformation
maps solutions of \eq{e1}-\eq{e2} into solutions, \emph{without
changing} the right-hand-sides of \eq{e3}-\eq{e5}:
\bel{transf}
 \mcE \to \mcE + 2\bar\lambda \varphi - |\lambda|^2 + \ii \mu\;, \quad
 \varphi \to \varphi - \lambda
 \;.
 \ee
This is easiest seen by noting, first, that both $f$ and  $d\mcE + 2
\bar\varphi d\varphi$ are left unchanged by \eq{transf}.

\section{\texorpdfstring{\boldmath$\mcE$}{E}-dominated ergosurfaces}
\label{SEdom}

 Suppose that $\Ef \ne \emptyset$ and that $ \mcE$ and
$\varphi$ are smooth in a neighborhood of $\Ef$. Let $z_0=\rho_0
+\ii \zeta_0 \in \Ef $, we can choose $\mu$ and $\lambda$ so that
the potentials transformed as in \eq{transf} satisfy
\bel{mlcond} \mcE (z_0)=0\;, \quad \varphi(z_0)=0
 \;.
\ee
Assume, first,
$$
Df(z_0)\ne 0
 \;.
$$
Performing a Taylor expansion of $\mcE$ and $\varphi$ at $z_0$ and
inserting into \eq{e1}-\eq{e2}, a {\sc Singular}~\cite{Singular}
 calculation (and, as a cross-check, a {\sc Maple} one) shows\footnote{See the
   {\sc Singular} file {\tt em1.in} and the {\sc Maple}
file {\tt em1.mw} at
\url{http://th.if.uj.edu.pl/~szybka/CS/}} that
either
\beal{Taylor1}
 &
\partial_z \varphi(z_0)= \partial_z \mcE (z_0) =0
 \;,
 &
 \\
 &
 0\ne \partial_\oz \mcE(z_0) = 4 \rho_0 \partial_z\partial_\oz \mcE (z_0)=  4 \rho_0\overline{\partial_z^2 \mcE} (z_0)
 \;,
 &
 \label{Taylor2}
 \\
 &
{\partial_z^2 \mcE} (z_0)\partial_z\partial_\oz \varphi
(z_0)={\partial_z^2 \varphi} (z_0)\partial_z\partial_\oz \mcE (z_0)
\;, &
  \label{Taylor3}
  \\
 &
{\partial_z^2 \mcE} (z_0)\overline{\partial_z^2 \varphi
}(z_0)=\overline{{\partial_z \partial_\oz \varphi}}
(z_0)\partial_z\partial_\oz \mcE (z_0) \;, &
  \label{Taylor4}
\eea
or that \eq{Taylor1}-\eq{Taylor4} is satisfied by the complex
conjugates of $(\mcE, \varphi)$. In the latter case the linear part
of the Taylor expansion of $(\mcE,\varphi)$ is a holomorphic
function of $z$, while it is anti-holomorphic in the former. In the
calculations proving smoothness across $\Ef \cap
 \{df \ne 0\}$ the equations \eq{Taylor3}-\eq{Taylor4} are not used.

Using \eq{Taylor2} in \eq{e5} one finds
\bel{aeq1} \lim_{z\to z_0} f^2\partial_z  \left(a+\frac \rho f\right) =
\lim_{z\to z_0} \partial_z  \left[f^2\partial_z  (a+\frac \rho f)\right]=
\lim_{z\to z_0} \partial_\oz  \left[f^2\partial_z  (a+\frac \rho f)\right]=0
 \;.
 \ee
 It follows as in the proof of Theorem~4.1 of \cite{CGMS} that the
 function $a+\rho/f$ is smooth across $\Ef  \cap
 \{df \ne 0\}$.

The same argument with $a-\rho/f$ instead of $a+\rho/f$ applies if
the complex conjugate solution is used.

 A similar calculation with \eq{e4} shows that
\bel{heq1} \lim_{z\to z_0} f^2 \partial_z \ln (|h/f|)= \lim_{z\to
z_0}
\partial_z (f^2 \partial_z  \ln (|h/f|))= \lim_{z\to
z_0}
\partial_\oz (f^2 \partial_z  \ln (|h/f|)) = 0\;.
 \ee
The remaining arguments of the proof of Theorem~4.1 of \cite{CGMS}
apply and we conclude that the metric \eq{2.1} extends smoothly
across $\Ef \cap
 \{df \ne 0\}$, and has Lorentzian signature in a neighborhood of
 this set.

Suppose, next,  that $f$ has a zero of higher order at $z_0\in \Ef$.
Since $\varphi$ enters quadratically in $f$ and in the
right-hand-sides of \eq{e4}-\eq{e5}, and through cubic terms in the
right-hand-sides of \eq{e1}-\eq{e2}, one would hope that $\varphi$
will only contribute to subleading terms in Taylor expansions of
those equations. But then the analysis of the leading-order behavior
of $f$ near $\Ef$ is reduced to the analysis already done
in~\cite{CGMS}, which would prove smoothness of the space-time
metric at the Ernst ergosurface without any provisos.

It turns out that this is not the case:  we shall see in the next
section that there exist leading-order  Taylor polynomials
satisfying the leading-order  equations for which the $\varphi$
terms are \emph{not} dominated by $\mcE$. Nevertheless, the argument
just given establishes that \emph{if} the $\varphi$ terms are
dominated by $\mcE$, then the analysis of \cite{CGMS} proves
smoothness of the metric across $\Ef$, and Theorem~\ref{TA} is
proved.

 \begin{Remark}
 \label{Redz}
Consider a $\mcE$-dominated zero  $z_0$ of $f$, after shifting
$\Im\mcE$ by a real constant we can assume that $\mcE(z_0)=0$. It
then follows from \cite[Proposition~5.1]{CGMS} that the order of the
zero of $\mcE$ at $z_0$ coincides with the order of the zero of $\Re
\mcE$.
 \end{Remark}

\section{Harrison--Neugebauer--Kramer transformations}
\label{SHarrison}

It is of interest to enquire what happens with Ernst ergosurfaces
under Neugebauer--Kramer
transformations~\cite[Equation~(34.8e)]{Exactsolutions2} (see
also~\cite{NeugebauerKramer69}) of $(\mcE,\varphi)$:
\begin{eqnarray}
\nonumber
 &
 \mcE'=\mcE(1-2\bar\gamma\varphi-\gamma\bar\gamma\mcE)^{-1},
 &
 \nonumber
 \\
 &
 \varphi'=(\varphi+\gamma\mcE)(1-2\bar\gamma\varphi-\gamma\bar\gamma\mcE)^{-1}
 \;.
 &
 \label{34.8e}
\end{eqnarray}
Under \eq{34.8e} $f$ is transformed to
\bel{fNG}  f
'=\frac{f}{|1-2\bar\gamma\varphi-\gamma\bar\gamma\mcE|^2}
 \;,
\ee
so that $\Ef$ is mapped into itself. The same remains of course true
under  Harrison~\cite{harr}
transformations~\cite[Equation~(34.12)]{Exactsolutions2}, which are
a special case of \eq{34.8e} when the initial $\varphi$ vanishes:
\begin{eqnarray}\label{34.12}
\mcE'=\mcE(1-\gamma\bar\gamma\mcE)^{-1},&&\varphi'=\gamma\mcE(1-\gamma\bar\gamma\mcE)^{-1}
 \;.
\end{eqnarray}

As a significant corollary of Theorem~\ref{TA}, we obtain

\begin{Corollary}
 \label{CHarrison} Let $(\mcE',\varphi')$ be obtained by a Harrison
transformation from a smooth solution $(\mcM,g)$ of the
\underline{vacuum} equations with a non-empty ergosurface, then the
conclusion of Theorem~\ref{TA} holds.
\end{Corollary}

\proof   As discussed in~\cite{CGMS}, the Ernst potential  $\mcE$ is
analytic near $\Ef$, hence has a zero of finite order. Clearly, the
order of zero of $|\varphi'|^2$ as defined by \eq{34.12} is higher
than the order of zero of $\mcE'$; the latter is the same as the
order of zero of $\Re \mcE'$ by the results in \cite{CGMS}.
\qed

Somewhat more generally, consider $p\in \Ef$, as explained above we
can always introduce a gauge so that $\varphi(p)=0$. In this gauge,
let $(\mcE',\varphi')$ be obtained by a Neugebauer--Kramer
transformation from a solution satisfying the hypotheses of
Theorem~\ref{TA} near $p$, then the conclusion of Theorem~\ref{TA}
holds near $p$ for the metric constructed by using
$(\mcE',\varphi')$. This follows immediately from \eq{34.8e}.

\section{Some remaining possibilities}
\label{Spartial}

It remains to consider the case where the $\varphi$ terms dominate
in $f$, and the case where all terms are of the same order. The
latter case will be referred to as \emph{balanced}.

\subsection{Balanced leading-order solutions with singular ergocircles}
\label{sScbl}

The simplest such possibility is $Df(z_0)=0$, $DDf(z_0)\neq 0$ and
$\mcE(z_0)=\varphi(z_0)=0$. It is easy to completely analyse the
first few leading-order  equations with the ansatz
\begin{eqnarray}
\label{Econs}
&\partial_z\mcE(z_0)=\partial_\oz\mcE(z_0)=\partial_z^2\mcE(z_0)=\partial_\oz^2\mcE(z_0)=0\;.&
\end{eqnarray}
A  {\sc Maple}--assisted calculation\footnote{See
the {\sc Maple} file {\tt em2.mw} at
\url{http://th.if.uj.edu.pl/~szybka/CS/}
} then shows that the leading-order equations do not introduce any
constraints on $\partial_z\varphi(z_0)$, and that if we set
$$
\alpha:=\partial_z\varphi(z_0)\ne 0\;,
$$
then one has
\begin{eqnarray}
& |\partial_\oz\varphi(z_0)|^2=|\alpha|^2\;,&\\\nonumber
&\partial_z\partial_\oz\mcE(z_0)=-4|\alpha|^2\;.&
\end{eqnarray}

Recall that \eq{e4}-\eq{e5} leads to the following equations for the
metric function $a$
\begin{eqnarray}
\label{eq:s1a}\frac{f^2}{\rho}\partial_z  (a+\frac \rho
f)&=&\underbrace{\left(\frac{\partial\mcE}{\partial
    z}+2\ovphi\frac{\partial\varphi}{\partial z}+\frac{f}{z+\oz}\right)}_{=:\hsone}
    \;,
    \\
\label{eq:s1b}\frac{f^2}{\rho}\partial_z  (a-\frac \rho
f)&=&\underbrace{-\left(\frac{\partial\omcE}{\partial
    z}+2\varphi\frac{\partial\ovphi}{\partial z}+\frac{f}{z+\oz}\right)}_{=:\hstwo}
    \;.
\end{eqnarray}
In the vacuum case it was shown that one out of $\hsone/f^2$ and
$\hstwo/f^2$ is smooth near $\{f=0,\; \rho \ne 0\}$, which then
implies smoothness of the ergosurface. (An identical analysis
applies to $\mcE$-dominated ergosurfaces.) So one can attempt to
repeat the argument here. Letting
$$
 r_0:=\sqrt{(\rho-\rho_0)^2+(\zeta-\zeta_0)^2}
 \;,
$$
the leading terms of $f$, $\hsone $, $\hstwo $ read
\begin{eqnarray}
 \nonumber
 \mcE&=&-4|\alpha z|^2+O(r_0^3)\;,
 \\
 \nonumber
\varphi&=&\alpha z+\bar\gamma \oz  +O(r_0^2) \;,
    \\
    f&=&-\alpha\gamma z^2+2|\alpha|^2 z\oz-\bar{\gamma}\bar{\alpha}\oz^2 +
O(r_0^3)
 \;,
 \\
 \nonumber
 \hsone &=&2\alpha(\gamma
z-\bar\alpha\oz)+O(r_0^2)\;,\\\nonumber \hstwo &=&-2\alpha(\gamma
z-\bar\alpha\oz)+O(r_0^2)\;,
\end{eqnarray}
where $\gamma=\overline{\partial_\oz\varphi}(z_0)$. Here, for
typesetting convenience, we used the symbol $z$ for $z-z_0$. Those
examples clearly lead to a singularity both in $\hsone /f^2$ and in
$\hstwo /f^2$, therefore a different strategy is needed.

Now,
\newcommand{\cz}{\bar z}%
$$
 f = |\alpha z - \bgamma \cz|^2 + (|\alpha|^2-|\gamma|^2) |z|^2 +
 O(r_0^3)
 \;,
$$
so that if $|\alpha|> |\gamma|$ we obtain an isolated zero of $f$,
an ``ergocircle".  More precisely, the intersection of the set where
$f$ vanishes with a neighborhood of $z_0$ will be $\{z_0\}$. This,
at any given value of $t$, corresponds to an isolated null orbit of
the isometry group of the metric generated by $\partial_\phi$
provided that the metric is non-singular there.

Still assuming $|\alpha|> |\gamma|$, we claim that the metric will
be singular at $z_0$. Indeed, adding \eq{eq:s1a} and \eq{eq:s1b} one
finds that $\partial a$ is uniformly bounded near $z_0$, hence $a$
can be extended by continuity to a Lipschitz continuous function
defined on a neighborhood of $z_0$. But then
$g(\partial_\phi,\partial_\phi)$ blows up as $\rho_0^2/f$ at $z_0$.

\subsection{Balanced solutions with radial {\texorpdfstring{\boldmath$\mcE_{2k}$}{E\_2k}}}

The solutions of Section~\ref{sScbl} are a special case of a family
of solutions in which the leading terms in $\mcE$ take the form
\begin{equation}\label{eq:mcE1}
\mcE_{2k} = \mu_1 e^{\ii \mu_0} \zzp^k \ozzp^k \;, \quad \mu_0\in
\R\;,\quad \mu_1\in \R^*\;.
\end{equation}
Let us write
\begin{equation}\label{eq:vp1}
\varphi_{k} = \sum_{m=0}^{k} \alpha_m \zzp^m \ozzp^{k-m},
\end{equation}
where all the $\alpha_m$'s do not vanish simultaneously.
Inserting~\eq{eq:mcE1}-\eq{eq:vp1} into \eq{e1}-\eq{e2} one obtains
\begin{equation}\label{eq:EM1}
(\mcE_{2k}+\omcE_{2k})\frac{\partial^2\mcE_{2k}}{\partial
\oz\partial
    z}-2\frac{\partial\mcE_{2k}}{\partial\oz}\frac{\partial\mcE_{2k}}{\partial
    z}=2\ovphi_{k}\left(\frac{\partial\varphi_{k}}{\partial\oz}\frac{\partial\mcE_{2k}}{\partial
    z}+\frac{\partial\varphi_{k}}{\partial z}\frac{\partial\mcE_{2k}}{\partial
    \oz}\right)-2\ovphi_{k}\varphi_{k}\frac{\partial^2\mcE_{2k}}{\partial\oz\partial z},\vspace{0.3cm}
\end{equation}
\begin{equation}\label{eq:EM2}
(\mcE_{2k}+\omcE_{2k})\frac{\partial^2\varphi_{k}}{\partial
\oz\partial
    z}-\left(\frac{\partial\varphi_{k}}{\partial\oz}\frac{\partial\mcE_{2k}}{\partial
    z}+\frac{\partial\varphi_{k}}{\partial z}\frac{\partial\mcE_{2k}}{\partial
    \oz}\right)=4\ovphi_{k}\frac{\partial\varphi_{k}}{\partial\oz}\frac{\partial\varphi_{k}}{\partial
    z}-2\ovphi_{k}\varphi_{k}\frac{\partial^2\varphi_{k}}{\partial\oz\partial z}.\vspace{0.3cm}
\end{equation}
The right-hand-side of  (\ref{eq:EM1}) vanishes, and the vanishing
of the left-hand-side implies \mbox{$\sin\mu_0=0\Longrightarrow
\mu_0=j\pi$}, where $j\in\mathbb{N}$. Changing $\mu_1$ to $-\mu_1$
if necessary we can without loss of generality assume $\mu_0=0$.
Setting $\alpha_i=0$ for $i<0$ or $i> k$, and working out the
coefficients of the terms \mbox{$\zzp^{k-1+l}\ozzp^{2k-1-l}$} in
(\ref{eq:EM2}) we obtain for $
 -k+1\leq l \leq 2k-1$
\begin{equation}
 \mu_1\alpha_l\left((k-l)^2+l^2\right)=-
\sum_{\begin{array}{c}{\scriptstyle
    -m+n+i=l}\\ {\scriptstyle 0\leq
    m,n,i \leq k}
\end{array}}
2\bar\alpha_m\alpha_n\alpha_i(k-i)(2n-i)\;.
\end{equation}
{We expect that a complete description of such solutions should be
possible (for example, it immediately follows for $2k-1>k$ (i.e.,
$k>1$) that $\bar \alpha_0 \alpha_k \alpha_{k-1}=0$),
but we have not attempted to do that.
Instead we list here all such leading-order solutions for $k=2$ and
$k=3$, as calculated\footnote{See the {\sc Maple} file {\tt em3.mw} at
\url{http://th.if.uj.edu.pl/~szybka/CS/}} using {\sc Maple}:}
\beaa
k=2\;,\ \mcE_4&=&-|\alpha|^2 |z|^4 :\quad \varphi_2= \alpha |z|^2
 \;, \quad \alpha \in \C^*\;,
  \\
 \ \mcE_4&=&-4|\alpha|^2 |z|^4 :\quad
  \varphi_2= \alpha z^2+ \ogamma   \oz^2
 \;, \quad \alpha,\gamma   \in \C^*\;,\ |\alpha|=|\gamma  |\;,
 \\
k=3\;,\ \mcE_6&=&-\frac 45|\alpha|^2 |z|^6 :\quad \varphi_3= \alpha
z|z|^2 \ \mbox{ or } \ \varphi_3= \alpha \oz|z|^2
 \;, \quad \alpha \in \C^*\;,
 \\
\mcE_6&=&-4|\alpha|^2 |z|^6:\quad \varphi_3= \alpha z^3+ \ogamma   \oz^3
 \;, \quad \alpha,\gamma   \in \C^*\;,\ |\alpha|=|\gamma  |\;.
\eeaa
As before, for typesetting convenience, we used the symbol $z$ for
$z-z_0$. (We have not included the solutions with $\varphi_k=0$, as
they are not balanced.)

The above suggests the following solutions, for all $k\ge 1$,
\beal{solRad2k}
  \mcE_{2k}&=&-4|\alpha|^2 |z|^{2k} :\;
 \varphi_k= \alpha z^k+ \ogamma   \oz^k
 \;, \; \alpha,\gamma   \in \C^*,\ |\alpha|=|\gamma  |,
 \\
 \label{soldRad4k}
  \mcE_{4k}&=&-|\alpha|^2 |z|^{4k} :\quad
 \varphi_{2k}= \alpha |z|^{2k}
 \;, \quad \alpha   \in \C^* \;,
 \\
 \mcE_{4k+2}&=&
-{\frac { 2\, k \left( k+1
 \right)  \left| \alpha \right|  ^{2}}{2\,k^{2}+2\,{k}+1}} |z|^{4k+2} :\nonumber\\
 && \varphi_{2k+1}= \alpha z|z|^{2k} \
\mbox{ or } \ \varphi_{2k+1}= \alpha \oz|z|^{2k}
 \;, \quad \alpha \in \C^*\;.
\eeal{soldRad4kp2}
Those can be  verified by a direct calculation.

Regularity of the metric can be established by showing that
$g_{\phi t}=-af$, $\ln g_{\zeta\zeta}=\ln g_{\rho\rho}=\ln
(hf^{-1})$, $g_{\phi\phi}=\left(\rho^2-(af)^2\right)/f$ are
smooth across $\{f=0,\rho>0\}$ and that $af$ does not vanish
whenever $f$ does. All solutions with leading-order  behavior
\eq{soldRad4k}, if any, have a zero of $f$ which is of order higher
than $4k$. Thus $f$ vanishes to higher order there, and any analysis
of the metric near $\{f=0\}$ requires knowledge of the higher-order
Taylor coefficients of $\mcE$ and $\varphi$ there.

On the other hand, the solution $\mcE_6=-4/5|\alpha|^2|z|^6 $,
$\varphi_3=\alpha z|z|^2$ leads to a singularity in the metric. (The
same is true for its conjugate pair, namely $\omcE$, $\bar\varphi$.)
For this solution we have, using \eq{e3}-\eq{e5},
\begin{eqnarray}
 \label{feq}
f&=&-\frac{1}{5}|\alpha|^2z^3\oz^3+\dots,\\
\frac{1}{h}\frac{\partial h}{\partial z}&=&-56\frac{\rho_0}{z^2}+\dots,\\
\frac{\partial a}{\partial
z}&=&25\frac{\rho_0}{|\alpha|^2z^4\oz^3}+\dots.
\end{eqnarray}
(\Eq{feq} shows that $f$ vanishes  at an isolated point in the
$(\rho,\zeta)$ plane, leading to again to an ergocircle.)
Integrating we obtain
\begin{eqnarray}
\ln (-h)&=&112\rho_0\frac{\rho-\rho_0}{(\rho-\rho_0)^2+(\zeta-\zeta_0)^2}+\dots,\\
a&=&\frac{-25}{3|\alpha|^2}\frac{\rho_0}{((\rho-\rho_0)^2+(\zeta-\zeta_0)^2)^3}+\dots,
\end{eqnarray}
hence
\begin{eqnarray}
af&=&\frac{5}{3}\rho_0+\dots,
 \\
\ln(hf^{-1})&=&112\rho_0\frac{\rho-\rho_0}{(\rho-\rho_0)^2
+(\zeta-\zeta_0)^2}
 \nonumber
 \\&&-\ln\left(\frac{1}{5}|\alpha|^2\left((\rho-\rho_0)^2+(\zeta-\zeta_0)^2\right)^3\right)+\dots,
 \\
g_{\phi\phi}&=&\frac{80}{9|\alpha|^2}\frac{\rho_0^2}{((\rho-\rho_0)^2+(\zeta-\zeta_0)^2)^3}+\dots.
\label{gphiphi}
\end{eqnarray}
Even though $af$ is regular at leading order,  the metric is
singular at the point $(\rho_0,\zeta_0)$. This is not merely a
coordinate singularity, since \eq{gphiphi} shows that the norm
$g_{\phi\phi}=g(\partial_\phi,\partial_\phi)$ of the
Killing vector $\partial_\phi$ is unbounded.

\subsection{\texorpdfstring{\boldmath$\varphi$}{phi}-dominated ergocircles}
 \label{Svde}

We consider now those solutions where $\varphi$ dominates in $f$. It
follows immediately from Theorem~\ref{Tdphi} below that such
solutions correspond to isolated points of $\{f=0\}$, hence to
ergocircles within the level sets of the coordinate $t$.

The simplest solutions in this class would have $\mcE$ vanishing
altogether, or vanishing to very high order. In this context,
symbolic algebra calculations%
\footnote{See the {\sc Singular} files {\tt em4a.in}, {\tt em4b.in} at
\url{http://th.if.uj.edu.pl/~szybka/CS/}}
show that there are no non-trivial solutions
such that
\begin{itemize}
\item $\varphi = O(|z-z_0|)$ with non-zero gradient at $z_0$, and
$\mcE=O(|z-z_0|^4)$,
\item $\varphi = O(|z-z_0|^2)$ with non-zero Hessian at $z_0$, and $\mcE=O(|z-z_0|^9).$
\end{itemize}
In other words  the assumption that $\varphi = O(|z-z_0|)$ and
$\mcE=O(|z-z_0|^4)$ implies  $\varphi = O(|z-z_0|^2)$; similarly
$\varphi = O(|z-z_0|^2)$ and $\mcE=O(|z-z_0|^9)$ implies $\varphi =
O(|z-z_0|^3)$. Those results require the analysis of the Taylor
series of $\varphi$ to higher order.

More systematically, let us assume that the leading-order  Taylor
polynomial $\varphi_k$ of $\varphi$ is of order $k$, with the
corresponding Taylor polynomial for $\mcE$ is of order $\ell$, while
$\Re \mcE=O(|\zz|^m)$. The following  shows that for both for
balanced
 and for $\varphi$-dominated solutions the order of $\mcE$
cannot be smaller than that of $|\varphi|^2$ (compare
Remark~\ref{Redz}):

\begin{Proposition}
 \label{Phighsame}
Suppose that $\mcE=O(|\zz|^\ell)$, $\varphi =
O(|\zz|^k)$, and $\Re \mcE=O(|\zz|^m)$ with $m \ge 2k$,
then
\bel{goodlk}
 \ell\ge 2k
  \;.
\ee
%
\end{Proposition}

 \proof Assume that $\ell<2k$, then inspection of \eq{e1} gives
$$
 \partial_z \mcE_\ell \partial_\oz \mcE_\ell=0
 \;.
$$
Since $\mcE_\ell$ is purely imaginary this reads $|d\mcE_\ell|^2=0$,
and the result follows.
 \qed

Clearly $m\ge \ell$   under the hypotheses of
Proposition~\ref{Phighsame}, so  \eq{goodlk} implies $m\ge \ell \ge
2k$. We conclude that at a zero which is balanced we must have
$m=\ell$; equivalently the order of $\mcE$ equals that of $\Re
\mcE$. The same is true for $\mcE$-dominated solutions by
Remark~\ref{Redz}. It follows that the hypothesis that $\varphi$
dominates in $f$ is equivalent to
\bel{kdom}
 2k < \ell
 \;.
\ee
Supposing that $f$ vanishes at $(\rho_0\;,\zeta_0)=z_0$, \eq{e2}
becomes
\begin{eqnarray}
 \ovphi_k\varphi_k L\varphi_k &=&
      2\ovphi_k\frac{\partial\varphi_k}{\partial
    \oz}\frac{\partial\varphi_k}{\partial
    z}
    +O(r_0^{k+\ell-2})
    +O(r_0^{3k-3})
    \;.
    \label{e2b}
\end{eqnarray}
By \eq{kdom} the second term can be absorbed into the first one.
Since the first derivatives part of $L$ contributes terms which
vanish faster than the second derivative ones, inspection of the
leading-order  terms leads to the equation
\bel{phiTq} \varphi_k 
 \Delta_2
 \varphi_k= 2 |d\varphi_k|^2 \quad \Longleftrightarrow
\quad 
\Delta_2
 \varphi_k^{-1}=0 \;,
 \ee
on the set $\{\varphi_k \ne 0\}$, where $\Delta_2$ is the Laplace
operator of the metric $d\rho^2+d\zeta^2 $. (Similarly, $(\mcE\equiv
0\;,\varphi)$ is a solution of \eq{e1}-\eq{e2} if and only if
$\Delta_3 \varphi^{-1}=0$, where $\Delta_3$ is the Laplace operator
of the metric $d\rho^2+d\zeta^2+\rho^2 d\phi^2$.)

We have the following:

\begin{Theorem}
\label{Tdphi} Homogeneous polynomial solutions of \eq{phiTq} are either
holomorphic or anti-holomorphic.
\end{Theorem}

\proof
Let $\varphi_k$ be a homogeneous polynomial of order $k$ solving \eq{phiTq},
conveniently parameterised as
\begin{equation}\label{eq:varphi1}
\varphi_k = \sum_{m=0}^k \alpha_m \zzp^m \ozzp^{k-m} \;.
\end{equation}
In complex notation the truncated Ernst--Maxwell equation \eq{phiTq}
reads
\begin{equation}\label{eq:varphi2}
\varphi_k\frac{\partial^2 \varphi_k}{\partial z\partial\oz} =
2\frac{\partial \varphi_k}{\partial z}\frac{\partial
\varphi_k}{\partial\oz} \;.
\end{equation}
Inserting (\ref{eq:varphi1}) into (\ref{eq:varphi2}) we obtain
\begin{equation}
\sum_{1\le m+j\le 2k-1} (k-m)(m-2j) \alpha_m\alpha_j
\zzp^{m+j-1}\ozzp^{2k-m-j-1}=0 \;.
\end{equation}
Hence, for $1\le \ell \le 2k-1$:
 \bel{eq:varphi3}
\sum_{m+j=\ell,\;m\leq k}(k-m)(m-2j)\alpha_m\alpha_j=0 \;. \ee
For $\ell\leq k$ this equation can be written in the form
\begin{equation}\label{eq:varphi4}
\sum_{m=0}^{\ell}(k-m)(3m-2\ell)\alpha_m\alpha_{\ell-m}=0\;.
\end{equation}
We consider $\ell\leq k$. For $\ell=1$ we have
$$(k+1)\alpha_0\alpha_1=0\;.$$
Assume, first, that $\alpha_0\neq 0$. Then $\alpha_1=0$, and for
$\ell=2$ we obtain
$$2(k+2)\alpha_0\alpha_2=0\;,$$
thus $\alpha_2=0$. More generally, if we assume for some $\ell_0$
that $\alpha_m=0$ for $0<m<\ell_0$ we have from (\ref{eq:varphi4})
$$\ell_0(k+\ell_0)\alpha_0\alpha_{\ell_0}=0 \quad\Longrightarrow\quad \alpha_{\ell_0}=0\;.$$
We can repeat this argument for $\ell=\ell_0+1$ and continue up to
$\ell=k$. Therefore, assumption $\alpha_0\neq 0$ leads to
$\alpha_m=0$ for $0<m\leq k$ and $\varphi_k$ is holomorphic.
Similarly, replacing above $\varphi_k$ with its complex conjugate
reveals that $\alpha_k\neq 0$ implies anti-holomorphicity of
$\varphi_k$. Note that for $k=1$ we are done.

Next, we assume $k\ge 2$ and we turn to the case $\alpha_0=0$, $\alpha_k=0$.
Again, we consider $\ell\leq k$. The equation with $\ell=1$ has
already been shown to be satisfied, but for $\ell=2$ we have
$$(k-1)\alpha_1^2=0\;,$$
thus $\alpha_1=0$ since  $k\ne 1$. The value of $\ell=3$ gives no
new conditions but for $\ell=4$
$$(k-2)\alpha_2^2=0\;,$$
thus $\alpha_2=0$.

More generally, let us assume that $\alpha_m=0$ for \mbox{$0\leq
m<m_0\leq k/2$},
then (\ref{eq:varphi4}) for $\ell=2m_0$ implies
$$(k-m_0)\alpha_{m_0}^2=0\;,$$
hence we have a contradiction. We conclude  that $\alpha_0=0$ implies $\alpha_m=0$ for $0\leq m\leq
k/2$.

The above result applied to the complex conjugate of $\varphi_k$  shows that
$\alpha_k=0$ implies $\alpha_m=0$ for $k/2\leq m < k$, as desired.
\qed

\subsubsection{\texorpdfstring{\boldmath$\varphi$}{phi}-dominated leading-order  solutions with singular
ergocircles} \label{sSFlos}

We continue our analysis of $\varphi$  of order $k\geq 1$, with the
leading term of $\mcE$   of order $2k+1$ or higher, so that  $f$ is
$O(r_0^{2k})$. (Note that some possibilities for $k=1$ and $k=2$
have already been eliminated at the beginning of
Section~\ref{Svde}.)
Since the Ernst--Maxwell equations are invariant under transformation
\mbox{$\varphi\rightarrow c\varphi$}, $\mcE\rightarrow \bar c
c\mcE$, where $c$ is a complex constant, we can without loss of
generality assume that the  Taylor development $\varphiT$ of
$\varphi$, as truncated at order $k+1$, takes the form
\begin{equation}\label{eq:vp1again}
\varphiT  = \zzp^k+\sum_{m=0}^{k+1} \alpha_m \zzp^m \ozzp^{k+1-m}
\;.
\end{equation}
Similarly, we have
\begin{equation}\label{eq:mcE1b}
\mcE_{2k+1} = \sum_{m=0}^{2k+1} \iota_m \zzp^m \ozzp^{2k+1-m} \;.
\end{equation}
The function $f$ takes the form
\begin{equation}\label{eq:f_lt}
f=-\zzp^k\ozzp^k+O(r_0^{2k+1}).
\end{equation}
The leading terms in the Ernst--Maxwell equations appear in order
$4k-1$ and $3k-1$, respectively
\begin{eqnarray}\label{eq:ep2}
\varphiT \frac{\partial^2\mcE_{2k+1}}{\partial z\partial
  \oz}&=&\frac{\partial\varphiT }{\partial z}\frac{\partial\mcE_{2k+1}}{\partial \oz}\;,\\
2\ovarphiT \left\{\varphiT \left(\frac{\partial^2 \varphiT
}{\partial
  z\partial\oz}+\frac{1}{2(z+\oz)}\frac{\partial \varphiT }{\partial z} \right)-
2\frac{\partial \varphiT }{\partial z}\frac{\partial \varphiT
}{\partial\oz}\right\}&=&\frac{\partial
\mcE_{2k+1}}{\partial\oz}\frac{\partial \varphiT }{\partial z} \;.\label{eq:vp2}
\end{eqnarray}
It follows from (\ref{eq:ep2}) that
\begin{equation}\label{eq:eoz}
\frac{\partial \mcE_{2k+1}}{\partial \oz}=\hat C(\oz)\varphiT \;,
\end{equation}
where $\hat C(\oz)$ is arbitrary function of $\oz$. However, we have
assumed that $\mcE$ has leading term of order $2k+1$. The comparison
of (\ref{eq:eoz}) with (\ref{eq:mcE1b}) gives
\begin{equation}\label{eq:eoz2}
\frac{\partial \mcE_{2k+1}}{\partial
\oz}=(k+1)\iota_k\zzp^k\ozzp^k\;,
\end{equation}
thus, $\iota_m=0$ for $m\neq k$ and $m\neq 2k+1$.

(Somewhat more generally,  an identical argument proves that if
 $\mcE=O(|\zz|^\ell)$ and $\varphi =
O(|\zz|^k)$, with $2k<\ell$, $\varphi$  holomorphic to leading
order, then there exists $c\in \C$ such that   $\mcE_\ell$ takes the
form $
 \mcE_\ell = c \zzp^k\ozzp^{\ell-k}$.)

The field equations imply
\begin{eqnarray}
\label{eq:k1n}\frac{f^2}{\rho} \partial_z \ln \left(\left|\frac{h}{f}\right|\right)=\hat\kappa
 \;,
\end{eqnarray}
where
\begin{eqnarray}
\label{eq:k1}\hat\kappa&:=&\frac{1}{2}\left(\left(\frac{\partial\omcE}{\partial
z}+2\varphi\frac{\partial\ovphi}{\partial
z}+\frac{2f}{z+\oz}\right)\left(\frac{\partial\mcE}{\partial
z}+2\ovphi\frac{\partial\varphi}{\partial
z}\right)\right.\nonumber\\&&+\left.\left(\frac{\partial\mcE}{\partial
z}+2\ovphi\frac{\partial\varphi}{\partial
z}+\frac{2f}{z+\oz}\right)\left(\frac{\partial\omcE}{\partial
z}+2\varphi\frac{\partial\ovphi}{\partial
z}\right)\right.\nonumber\\&&-\left.4\frac{\partial\ovphi}{\partial
z}\frac{\partial\varphi}{\partial
z}\left(\mcE+\omcE+2\ovphi\varphi\right)\right) \;,\phantom{XXXX}
\end{eqnarray}
and recall that the functions $\hsone $ and $\hstwo $ have been
defined in \eq{eq:s1a}-\eq{eq:s1b}. We are going to show that if the
conditions mentioned at the beginning of this section hold, then
(\ref{eq:vp2}), (\ref{eq:ep2}) imply that
$$
 \hstwo =d\hstwo =\dots=d^{2k}\hstwo =0
$$
and
$$\hat\kappa=d\hat\kappa=\dots=d^{4k-2}\hat\kappa=0
$$
on
$E_{\mcE,\varphi}$    but $d^{4k-1}\hat\kappa=0$ there only for
special solutions.

Inserting  (\ref{eq:vp1again}) and (\ref{eq:eoz2}) into
(\ref{eq:vp2}) gives
\begin{eqnarray} \nonumber
&&\sum_{m=0}^{k-1}(k+1-m)(m-2k)\alpha_m\zzp^{k+m-1}\ozzp^{k-m}\\
 \label{eq:sum2}
&& -k\left(\alpha_k+\frac{k+1}{2}\iota_k-\frac{1}{4\rho_0}\right)\zzp^{2k-1}
 =0\;.
\end{eqnarray}
The comparison of the coefficients in front of powers of $\zzp$ and
$\ozzp$ allows us to read off that $\alpha_m=0$ for $m=0,
\ldots,k-1$. Moreover,
$$
 \alpha_k+\iota_k(k+1)/2=\frac{1}{4\rho_0}
$$
and there are no restrictions in the leading order on
$\alpha_{k+1}$, $\iota_{2k+1}$.
Hence
\beaa
  \varphiT  &=& \zzp^k+  \alpha_k \zzp^k \ozzp + \alpha_{k+1}
\zzp^{k+1} \;,
 \\
    \mcE_{2k+1}
    &=&
     \iota_k\zzp^k\ozzp^{k+1}\;.
\eeaa
%
 Keeping this result in mind, we write down the leading terms of
$\hstwo $:
\begin{eqnarray}
 \nonumber
 \hstwo
 &=&
 -\frac{\partial \omcE_{2k+1}}{\partial z}-2\varphiT
\left(\frac{\partial \ovarphiT }{\partial
  z}-\frac 12 \frac{\ovarphiT }{z+\oz}\right)+O(r_0^{2k+1})
  \\
  \nonumber
  &=&-2\left(\sum_{m=0}^k(k+1-m)\bar\alpha_m\ozzp^m \zzp^{2k-m}\right.\\&&\nonumber\left.+\Big(\frac{k+1}{2}\bar\iota_k
  -\frac{1}{4\rho_0}\Big)
  \ozzp^{k}\zzp^{k}\right)+O(r_0^{2k+1})
  \nonumber
  \\
  &=&
  O(r_0^{2k+1}).\label{eq:sigma_II_ex}
\end{eqnarray}
Therefore, $\hstwo $ is at least $O(r_0^{2k+1})$. Moreover, it
follows from the identity
\begin{equation}
-2\frac{\partial f}{\partial z}=\hsone -\hstwo -\frac{2f}{z+\oz},
\end{equation}
that $\hsone $ is $O(r_0^{2k-1})$ \emph{but not better},
because it has to compensate for the lowest terms of $\partial_z f$,
see (\ref{eq:f_lt}).

Now, we turn to $\hat\kappa$. Firstly, we rewrite (\ref{eq:k1}) in
terms of $\hsone $, $\hstwo $
\begin{equation}
\label{eq:k1a} \hat\kappa=-\hsone \hstwo
-\frac{f^2}{(z+\oz)^2}+4\frac{\partial\bar\varphi}{\partial
z}\frac{\partial\varphi}{\partial z}f\;.
\end{equation}
It follows from our previous results and  (\ref{eq:f_lt}) that
\begin{equation}
\hat\kappa=-\left(\frac{1}{\rho_0}-2(k+1)\bar\iota_k\right)k\zzp^{2k-1}\ozzp^{2k}+O(r_0^{4k}).
\end{equation}
Therefore, $\hat\kappa$ is only $ O(r_0^{4k-1})$ for any
$$
 \iota_k\neq
 (2(k+1)\rho_0)^{-1}
$$
and any solution with the above leading-order  behavior, if it
exists, will lead to a singular space-time metric (note, however,
that this could be a coordinate singularity).

On the other hand if $\iota_k=(2(k+1)\rho_0)^{-1}$ then $\alpha_k=0$
and $\varphi$ is holomorphic also in the order $k+1$. For such
solutions $\hat\kappa$ is at least $O(r_0^{4k})$, which is
\emph{not} incompatible in an \emph{obvious} way with smoothness of
the space-time metric at the ergosurface.

\section{Concluding remarks}

Our results are far from satisfactory, with the following questions
open:

\begin{enumerate}
\item Which ``solutions at leading order", as constructed above using Taylor series
expansions (whether balanced,   $\varphi$- or $\mcE$- dominated),
 \emph{do arise} from real solutions of
the Ernst--Maxwell equations which are smooth across the zero-level
set of $f$? Here we mean that the associated harmonic map is smooth,
without (in a first step) requesting that the associated space-time
metric be smooth as well. The non-existence results mentioned at the
beginning of section~\ref{Svde} are instructive: there \emph{do}
exist Taylor polynomials solving the leading-order  equations with
$\varphi = O(|z-z_0|)$ with non-zero gradient at $z_0$ and with,
say, $\mcE=0$, and one has to go a few orders more in the Taylor
series to show that the coefficients of the leading-order  Taylor
polynomial are all zero. The same mechanism applies to leading-order
solutions with $\varphi = O(|z-z_0|^2)$ with non-zero Hessian at
$z_0$.

\item Can one exhaustively describe the balanced leading-order  solutions?  The
question seems hard. There does not seem, however, to be any good
reason to invest a lot of energy therein as long as the previous
question remains open.
\end{enumerate}

{\sc Acknowledgements:} Part of this work was done when the first
author was visiting the Albert Einstein Institute, Golm. We are also
grateful to Jena University for hospitality. Useful conversations
with Laurent V\'eron are acknowledged.

\bibliographystyle{amsplain}

\bibliography{
../../references/reffile,%
../../references/newbiblio,%
../../references/bibl,%
../../references/howard,%
../../references/myGR,%
../../references/newbib,%
../../references/Energy,%
../../references/netbiblio,%
../../references/PDE}

\end{document}